\documentclass[longbibliography,aps,prb,preprint,superscriptaddress]{revtex4-2}

\usepackage{pdfpages}
\makeatletter
\AtBeginDocument{\let\LS@rot\@undefined}
\makeatother
\usepackage[latin9]{inputenc}

\usepackage{wrapfig}
\usepackage{units}
\usepackage{amsbsy}
\usepackage{amstext}
\usepackage{amsmath}
\usepackage{graphicx}
\usepackage{xcolor}
\usepackage{times}
\usepackage{float}
\usepackage{setspace}

\usepackage{hyperref}

\usepackage{amssymb}

\begin{document}


\title{Universality of the {\bf q}=1/2 Orbital Magnetism in the Pseudogap Phase of the High-$T_c$ superconductor  $\rm YBa_{2}Cu_{3}O_{6+x}$ }




\author{Dalila Bounoua}
\affiliation{Universit{\'e} Paris-Saclay, CNRS, CEA, Laboratoire L{\'e}on  Brillouin, 91191, Gif-sur-Yvette, France}
\author{Yvan Sidis}
\affiliation{Universit{\'e} Paris-Saclay, CNRS, CEA, Laboratoire L{\'e}on  Brillouin, 91191, Gif-sur-Yvette, France}
\author{Martin Boehm}
\affiliation{Institut Laue-Langevin, 71 avenue des Martyrs, Grenoble 38042, France}
\author{Paul Steffens}
\affiliation{Institut Laue-Langevin, 71 avenue des Martyrs, Grenoble 38042, France}
\author{Toshinao Loew}
\affiliation{Max Planck Institute for Solid State  Research, Heisenbergstrasse 1, 70569 Stuttgart, Germany}
\author{Lin Shan Guo}
\affiliation{Shanghai Jiao Tong University,  800 Dong Chuan Road, Shanghai 200240, China}
\author{Jun Qian}
\affiliation{Shanghai Jiao Tong University,  800 Dong Chuan Road, Shanghai 200240, China}
\author{Xin Yao}
\affiliation{Shanghai Jiao Tong University,  800 Dong Chuan Road, Shanghai 200240, China}
\author{Philippe Bourges}
\affiliation{Universit{\'e} Paris-Saclay, CNRS, CEA, Laboratoire L{\'e}on  Brillouin, 91191, Gif-sur-Yvette, France}






\begin{abstract}

\noindent  

Several decades of debate have centered around the nature of the enigmatic pseudo-gap state in high temperature superconducting copper oxides. Recently, we reported polarized neutron diffraction measurements that suggested the existence of a magnetic texture bound to the pseudo-gap phase [Bounoua, {\it et al}. Communications Physics 5, 268 (2022)]. Such a magnetic texture is likely to involve the spontaneous appearance of loop currents within the CuO$_2$ unit cells, which give birth to complex correlated patterns. In the underdoped ${\rm YBa_{2}Cu_{3}O_{6.6}}$, the magnetic structure factor of such an orbital magnetic texture gives rise to two distinct magnetic responses at {\bf q}=0 and {\bf q}=1/2. As this pattern alters the lattice translation invariance,  such a state of matter could contribute to an instability of the Fermi surface.  Here, we report polarized neutron scattering measurements on a nearly optimally doped high quality single crystal of ${\rm YBa_{2}Cu_{3}O_{6.9}}$ that exhibits the same {\bf q}=1/2 magnetism  and a weakly overdoped ${\rm YBa_{2}Cu_{3}O_{7}}$ sample where this signal is no longer sizeable. The in-plane and out-of-plane magnetic neutron scattering intensities in ${\rm YBa_{2}Cu_{3}O_{6.9}}$ (at {\bf q}=1/2) and ${\rm YBa_{2}Cu_{3}O_{6.85}}$ (at {\bf q}=0), reported previously, display the same temperature dependent hallmarks. The magnitudes of both {\bf q}=0 and {\bf q}=1/2 magnetic signals further exhibit the same trends upon doping in ${\rm YBa_{2}Cu_{3}O_{6+x}}$, confirming that they are likely intertwined.

\end{abstract}

\maketitle

\clearpage

\section{\label{Intro} Introduction}

The generic phase diagram of hole doped cuprate superconductors shows various electronic instabilities  \cite{Keimer15,Proust19,Varma20}. Beyond unconventional superconductivity, evidence for (spin, charge and pair) density waves were reported, in addition to intra-unit-cell orders breaking local Ising symmetries (time reversal, parity, rotation). 
The pairing mechanism that permits superconductivity in copper oxides remains unknown, although a mechanism based on magnetic interactions is generally favoured. For instance, the correspondence between the changes of the charge-transfer-gap (controlling the magnetic superexchange interaction) and the Cooper pair density demonstrates a strong and compelling correlation, hence perhaps constitutes the clearest evidence yet of a mechanism that underlies superconductivity in cuprates\cite{Mahoney22}. 

At present, there is a lack of consensus about the pseudogap  state of matter that appears below a temperature T* and from which the superconductivity emerges in a wide range of its phase diagram.  In the underdoped regime, the $d$-wave superconducting (SC) state competes with an incipient bi-axial charge density wave (CDW). Both states appear deep inside the so-called pseudogap (PG) phase, which creates gaps in extended portions of the Fermi surface and gives rise to Fermi arcs only. Concomitantly, an intra-unit-cell (IUC) {\bf q}=0 antiferromagnetism develops with the PG state  \cite{Fauque06,Mook08,Baledent11,deAlmeida12,Mangin15,Tang18,bourges2021-online}, without giving a net magnetization in the CuO$_2$ plaquette that would be expected for a ({\bf q}=0)  ferromagnetism. This state is usually associated with magneto-electric loop currents (LC) running through copper and oxygen orbitals, and whose order parameter can be conveniently described by a polar anapole vector \cite{Varma20,bourges2021-online}. 
 The observed magnetism then corresponds to orbital moments, a picture supported by the observation of an unusual form factor  \cite{deAlmeida12}. 
It breaks both time and parity symmetries, but preserves the lattice translation invariance \cite{Varma06,Varma20,Scheurer18,Sarkar19}.  Supporting these observations from polarized neutron measurements and muon spin spectroscopy confirms the existence of a slowly fluctuating magnetism in the PG state  \cite{zhang2018discovery}. The corresponding magnetic fluctuations are characterized by a finite timescale that does not evolve with temperature \cite{Zhu2021}.  An interesting characteristic of the IUC orbital magnetism is that it survives upon dimensional confinement, and its correlation lengths become short-ranged, as  observed in the quasi-1D ${\rm (Sr,Ca)_{14}Cu_{24}O_{41}}$ two leg-ladder cuprates \cite{Bounoua20}, or in ${\rm La_{2-x}Sr_{x}CuO_{4}}$ \cite{Baledent10} due to charge segregation (leading to charge stripes). However, it is worth emphasizing that the breaking of the discrete symmetries cannot alone open the PG \cite{Chatterjee17}.

Recently, polarized neutron diffraction studies uncovered the existence of a novel form of magnetism,  labelled {\bf q}=1/2, hidden in the pseudogap phase of superconducting cuprates \cite{Bounoua22}. In two distinct ${\rm YBa_{2}Cu_{3}O_{6.6}}$ underdoped samples,  a static short-range magnetic response is observed at the planar wave vector (0.5,0). Gathering all the information available so far, one can state that the {\bf q}=1/2 magnetism belongs to the {\rm CuO$_2$} planes and is observed at both (0.5,0) and (0,0.5) in a detwinned sample \cite{Bounoua22}. Breaking the lattice symmetry, the novel magnetism with a propagation vector $(\pi, 0)\equiv (0, \pi)$ can then be described by a 2x2 larger unit cell, which is inconsistent with known spin structures in cuprates, rather suggesting an orbital magnetism related to loop currents. Similarly to the {\bf q}=0 IUC magnetism, the  {\bf q}=1/2 magnetic signal  typically sets in at the PG onset temperature on cooling down from room temperature, suggesting that both magnetic responses are intertwined and are directly related to the PG physics. The deduced magnetic moment of the  {\bf q}=1/2 magnetism is further found to point predominantly along the c-axis \cite{Bounoua22}, in contrast to the {\bf q}=0 IUC magnetism whose ordered moment exhibits a systematic tilt of about 45$^\circ$ from the c-axis  \cite{Tang18}. 
Even though there is no definitive explanation for the origin of the tilt \cite{bourges2021-online}, the magnetic response was found to reorient from fully out-of-plane to tilted on cooling in ${\rm YBa_{2}Cu_{3}O_{6.85}}$\cite{Mangin15} supporting a possible scenario of an evolution from classical (above T*) to quantum LC states \cite{He12}, the classical LC state being characterized by the existence of a magnetic moment strictly perpendicular to the circulating loop currents. The out-of-plane c-polarized magnetic response favors interpretations related to LC running within the {\rm CuO$_2$} planes, naturally leading to orbital magnetic moments pointing along the c-axis. One can nevertheless wonder why the {\bf q}=1/2 magnetism should not exhibit an in-plane component as does the {\bf q}=0 IUC magnetism if both features were intertwined. 

The {\bf q}=1/2 magnetism remains at short range at all temperatures with a correlation length of about 25 \AA\ range, corresponding to the formation of clusters of 5-6 unit cells within the  {\rm CuO$_2$} planes \cite{Bounoua22}. Together with the previous reports of the {\bf q}=0 IUC magnetism in cuprates  \cite{Fauque06,Tang18,bourges2021-online}  (that develops at longer distance), the novel magnetism at $(\pi,0)$ could actually belong to a unique complex bi-axial magnetic texture of the {\rm CuO$_2$} unit cells hosting loop currents/anapoles \cite{Bounoua22}. This picture is very similar to the case of charge density instabilities where a {\bf q}=0  nematic charge order co-exists with the {\bf q}{$\neq$}0 charge density wave with shorter correlation lengths \cite{Fujita14} transposed in the orbital magnetism sector. While various magnetic patterns could account for the magnetic diffraction data available so far, one can for instance visualize such a magnetic texture as a supercell made of 4 large domains with {\bf q}=0 LC order domains (where loop current patterns/anapoles rotate at 90$^\circ$ from one domain to the other) and, at their junction, a bubble of intertwined 2x2 loop currents related to the local $(\pi, 0)$ magnetic response \cite{Bounoua22}. In both ${\rm YBa_{2}Cu_{3}O_{6.6}}$ samples, the {\bf q}=1/2  magnetic signal is weakly correlated along the c-axis, perpendicular to the CuO$_2$ planes, with a maximum structure factor centered at $L$=0 \cite{Bounoua22}.  If its observation is ubiquitous in the PG phase, the {\bf q}=1/2 magnetism is likely important for the understanding of the instrinsic nature of the PG state since it breaks the lattice translational symmetry  and  develops at the PG onset temperature. However, it was only reported for a single hole doping level, so far.

Here, using polarized neutron diffraction, we investigated the existence of  the {\bf q}=1/2 magnetism in two high quality single crystals at higher doping levels than in the former study (p$\simeq$ 0.11 for ${\rm YBa_{2}Cu_{3}O_{6.6}}$ \cite{Bounoua22}). In particular, we report the existence of the {\bf q}=1/2 magnetism in a nearly optimally doped  ${\rm YBa_{2}Cu_{3}O_{6.9}}$ single crystal, with a hole-doping of p$\simeq$ 0.16, where the {\bf q}=1/2 magnetism displays the same temperature dependence as  the {\bf q}=0 IUC magnetism reported for a similar doping \cite{Mangin15}. Compared to the ${\rm YBa_{2}Cu_{3}O_{6.6}}$ samples, the {\bf q}=1/2 magnetism is reduced in amplitude upon increasing doping in a similar way to  the {\bf q}=0 IUC magnetism. Interestingly, its c-axis dependence exhibits a complex behaviour with a structured magnetic response along  {\bf Q}=$(0.5,0,L)$, giving rise to two well-defined peaks at $L$=0 and $L$=1/2, suggesting a magnetic doubling of the unit cell in the stacking direction of CuO$_2$ planes along the c-axis. The magnetic scattering associated with the $L$=1/2 peak is found to be mainly related to the in-plane component in contrast to the $L$=0 peak reported so far  \cite{Bounoua22}, which arises predominantly from the out-of-plane magnetic component. The L-dependent orientation of the magnetic response indicates that the in-plane and out-of-plane magnetic components at {\bf q}=1/2 occur at different $L$ positions along {\bf Q}=$(0.5,0,L)$, in contrast with the {\bf q}=0 IUC magnetic response for which both types of components are superimposed in momentum space. 

\section{\label{Exp} Experimental details}
\subsubsection{\label{Polar} Polarized neutron scattering experiment}

The polarized neutron measurements were performed on two different neutron Triple Axis Spectrometers (TAS) with different momentum and energy resolutions. In both cases, the measurements were carried out in elastic conditions with the final neutron wavevector (k$_f$) equal to the incident one  (k$_i$). On the instrument 4F1 located at the LLB-Orph\'ee reactor in Saclay (France), the neutron wavevector was k$_f$=2.57 \AA$^{-1}$; a Pyrolitic Graphite (PG) filter installed on the incident beam was used to remove higher order harmonics. A double PG monochromator was used to select the incident neutron wavelength. The beam was polarized using a bender  supermirror located between the second monochromator and  the sample position and the final polarization state was analysed with an Heusler crystal Cu$_2$MnAl using the (1,1,1) Bragg reflection.  On Thales located at Institut Laue Langevin (ILL) in Grenoble (France), the neutron wavevector was much lower, k$_f$=1.5 \AA$^{-1}$, yielding a better energy resolution.  A velocity selector in the incoming and a Be filter in the outgoing beam were used to remove higher order harmonics.  The incoming and outgoing beam polarizations were realized using Heusler crystals. 

Both TAS instruments were equipped with longitudinal XYZ polarization analysis  (XYZ-PA) (see \cite{Bounoua22}), a powerful technique to selectively probe and disentangle the magnetic response from the nuclear one with no assumptions on the background. Both spin-flip (SF) and non-spin-flip (NSF) scans were done in order to crosscheck the absence of nuclear scattering at magnetic positions. According to conventional notations, $\bf X$ stands for the direction of the neutron polarization  parallel to the transferred momentum $\bf Q$, $\bf Y$ and $\bf Z$ polarization directions are both perpendicular to  $\bf Q$. $\bf Y$  is the in-plane orthogonal direction and $\bf Z$ is perpendicular to the scattering plane. The spin-flip, $SF_{X,Y,Z}$, and non-spin-flip, $NSF_{X,Y,Z}$, intensities for the different neutron polarizations were measured in order to be able to perform the XYZ polarization analysis. Throughout the article, the total scattered magnetic intensity,  $I_{mag}$, is extracted such as (See \cite{Bounoua20,Bounoua22} for more details about the analysis of the neutron data):

\begin{equation}
I_{mag}= 2\,SF_{\bf X}-{SF}_{\bf Y} - {SF}_{\bf Z}
 \label{Imag}
\end{equation}

The flipping ratios FR=$(\frac{NSF_{X,Y,Z}}{SF_{X,Y,Z}})$ on both instruments were measured on the sample Bragg peaks as well as on a quartz speciment. They were found to be homogeneous along $\bf X, Y$ and $\bf Z$ polarizations, such as (FR=15) on 4F1 TAS and (FR=40) on Thales TAS (see Supplemental Material  \cite{SuppMat} for more details). On Thales, the sample was mounted in a  spherical polarization analysis device, CryoPAD\cite{Cryopad}, with zero magnetic field at the sample chamber that allows one to perform XYZ-PA even in the superconducting state without risks of neutron spin depolarization. On 4F1, the neutron polarization at the sample position was controlled using MuPad\cite{Mupad}, another type of spherical polarization analysis device where the sample is installed inside a zero magnetic field chamber.  As a result, the polarisation is very homogeneous with tiny variations of the flipping ratio upon turning the polarization. Therefore, corrections from imperfect neutron polarization have negligible effects in Eq. \ref{Imag} (see an example in Supplemental Material  \cite{SuppMat}).

\subsubsection{\label{sample} Sample preparation and characterization }

\begin{figure}
 \begin{centering}
 \includegraphics[width=15cm]{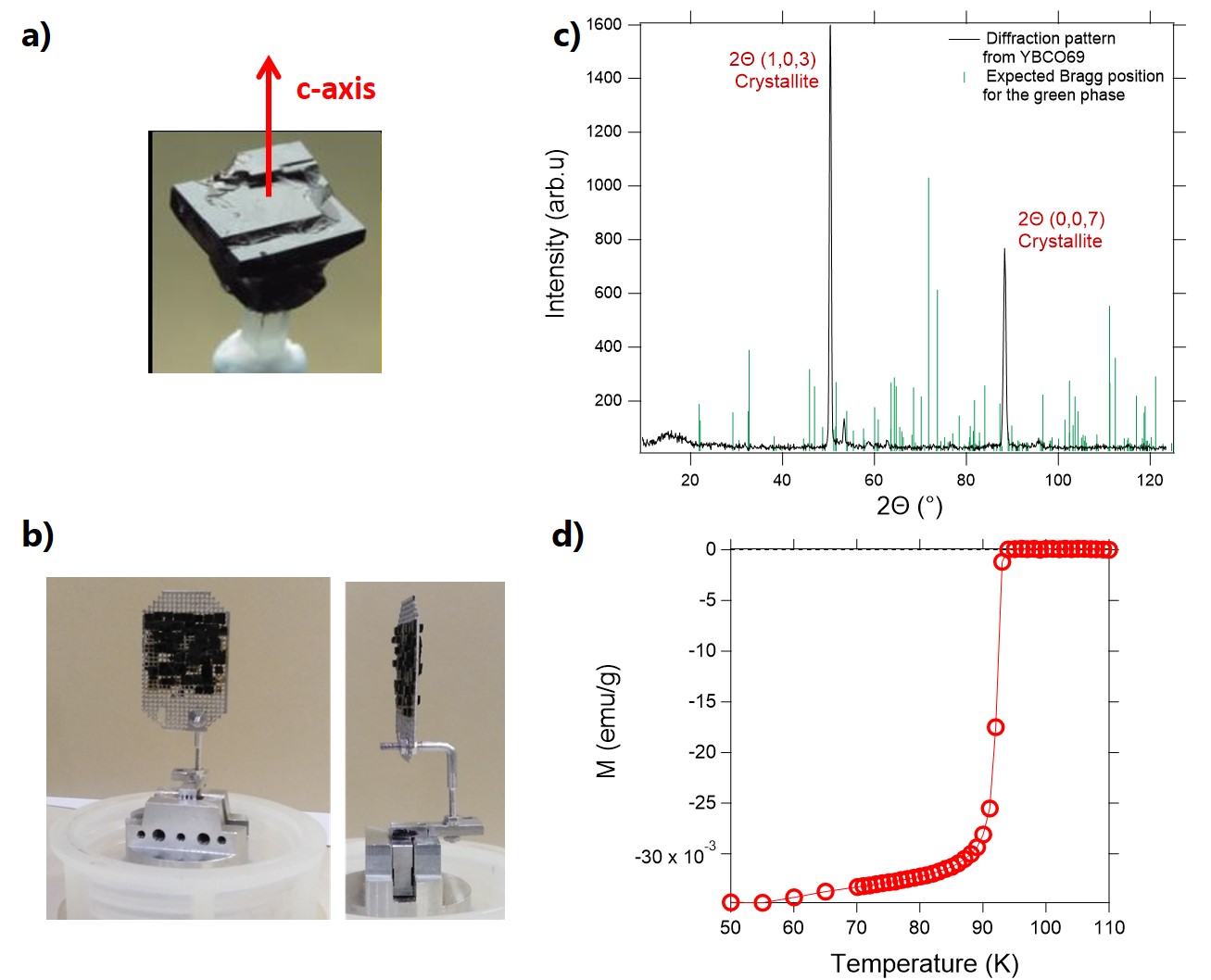} 
 \par\end{centering}
 \clearpage
\caption{\label{Fig1} \textbf{YBCO69 and YBCO7 samples}: \textbf{(a)} Single crystal of YBCO69 and \textbf{(b)} co-aligned detwinned single crystals of YBCO7 used for the polarized neutron study. \textbf{(c)} Diffraction pattern recorded at $k_i$=2.69 \AA$^{-1}$ on the 3T1 diffractometer at LLB-Orph\'ee for YBCO69 aligned in an arbitrary orientation to avoid the main Bragg peaks. The green lines represent the expected powder diffraction from the so-called green phase ${\rm Y_{2}BaCuO_{5}}$. None of the lines from the green phase are present in the YBCO69 diffractogram where the two peaks correspond to scattering from tails of (1,0,3) and (0,0,7)  Bragg reflections of YBCO69. \textbf{(d)} Field cooled magnetization versus temperature data recorded for YBCO69 under an applied field of 10 Oe showing the occurrence of the superconducting transition around $T_c$=91.9K. 
 }
\end{figure}

Two different ${\rm YBa_{2}Cu_{3}O_{6+x}}$ (YBCO) samples near optimal hole doping were studied (Fig. \ref{Fig1}). Both samples exhibit the ortho-I orthorhombic structure where the Cu-O chains fully align along the $b$ axis  for x$\ge$0.85  \cite{Zimmermann03}. The first sample is a large  ${\rm YBa_{2}Cu_{3}O_{6.9}}$ single crystal, hereafter sample YBCO69 (Fig. \ref{Fig1}.a). It was grown at Shanghai Jiao Tong University using a polythermal method of top-seeded solution-growth to get large  ${\rm YBa_{2}Cu_{3}O_{6+x}}$ single crystals \cite{Xiang16}. The YBCO69 sample was never studied with polarized neutron diffraction before, however YBCO single crystals grown with this growth technique have been already studied with inelastic neutron scattering \cite{ShiliangLi08} and quantum oscillations \cite{Singleton10} among other technique. The sample is twinned with a mass of $\simeq$7 g and contains no impurity such as the so-called green phase ${\rm Y_{2}BaCuO_{5}}$ (Fig. \ref{Fig1}.c.) or cupric oxide CuO (see Supplemental Material  \cite{SuppMat}).  The as-grown single crystal was annealed for 18 days at 400$^{\circ}$C, followed by 40 days heat treatment at 520$^{\circ}$C under oxygen flow \cite{Gao06} to ensure the diffusion of oxygen into the bulk, then quenched in liquid nitrogen. After annealing, the sample exhibits a sharp superconducting transition temperature at $T_c$ = 91.9 K (Fig. \ref{Fig1}.d).  The c-axis parameter was further measured using neutron diffraction. Using the standard determination of hole doping \cite{Liang06}, one obtains p$\simeq$ 0.16.  

The second sample is an array of detwinned  ${\rm YBa_{2}Cu_{3}O_{7}}$ crystals (hereafter sample YBCO7 shown on Fig.\ref{Fig1}.b) , grown at the Max Planck Institute for Solid State Research in Stuttgart. This sample was previously used for an inelastic neutron scattering experiment to study phonon anomalies \cite{Raichle11}.
 The detwinning ratio of the array is 86\%.  It has a mass of 2.6 g and exhibits a superconducting transition temperature at 90K  \cite{Raichle11}, leading to p$\simeq$ 0.19. 

The YBCO69 sample was studied on both the 4F1 and Thales instruments in  the (1,0,0)/(0,0,1) scattering plane such that wave-vectors of the form $(H,0,L)$  were accessible. The YBCO7 sample (only measured on 4F1) was instead mounted in the (1,0,0)/(0,1,0) scattering plane to study wave-vectors of the form $(H,K,0)$. In the following, the wave-vectors are indexed in reduced lattice units (r.l.u.), in units of $(\frac{2\pi}{a}, \frac{2\pi}{b},\frac{2\pi}{c})$ where $a$, $b$ and $c$ stand for the lattice parameters of the sample.

\section{\label{results} Experimental results}
\subsubsection{\label{Order} Short-range magnetic correlations}

\begin{figure}
 \begin{centering}
 \includegraphics[width=15cm]{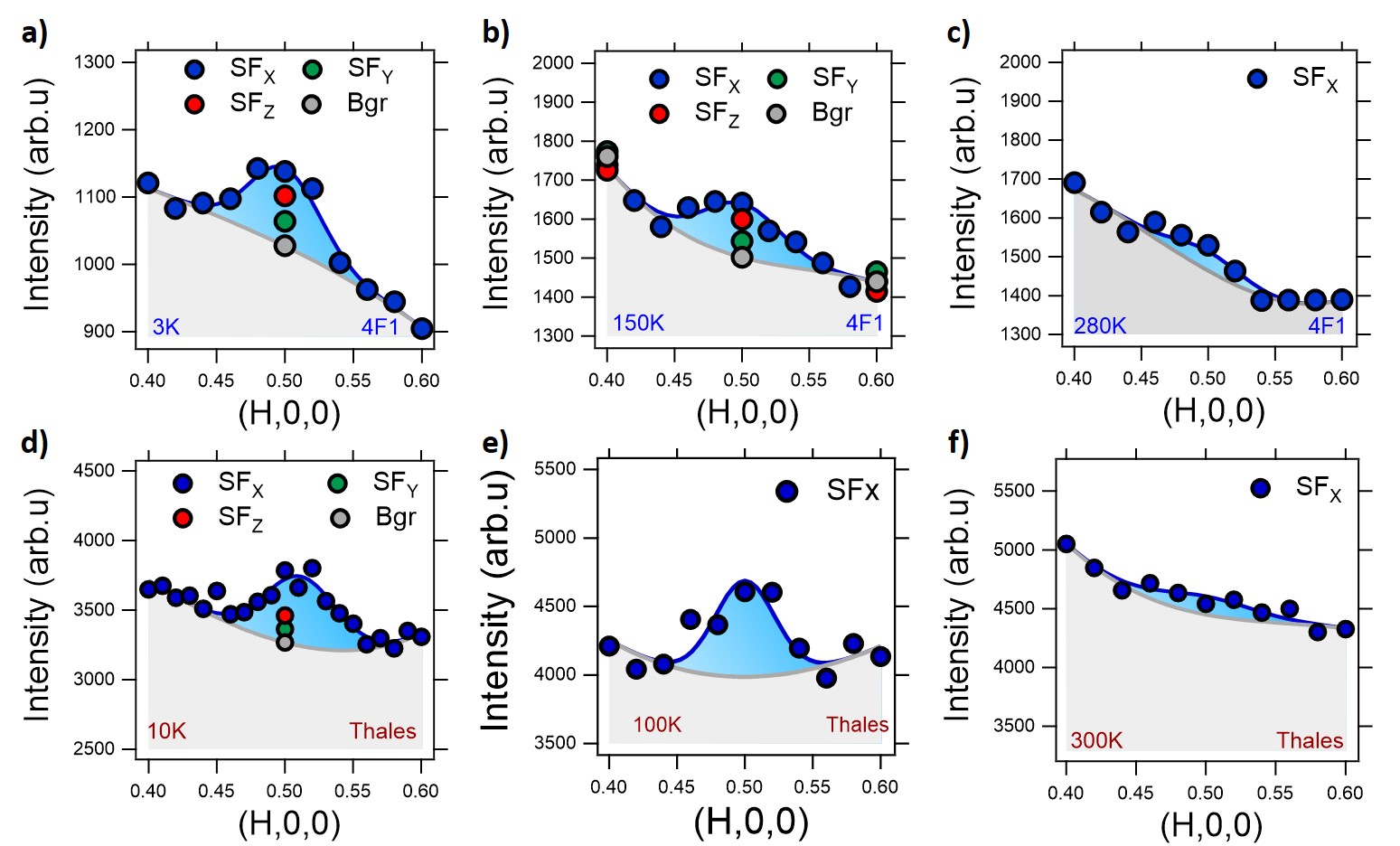} 
 \par\end{centering}
 \clearpage
\caption{\label{Fig2} \textbf{{\bf q}=1/2 orbital magnetism in twinned $\rm YBa_{2}Cu_{3}O_{6.9}$ (YBCO69)}: Scans along (H,0,0) in the spin-flip channel  $SF_{X,Y,Z}$. The magnetic intensity appears as a commensurate peak centered at (0.5,0,0) in \textbf{(a-f)}. Data in \textbf{(a-c)} were measured on the 4F1 TAS at 3, 150, and 280K, respectively. Data in \textbf{(d-f)} were measured on the Thales TAS at 10, 100, and 300K, respectively. The sample was aligned in the (1,0,0)/(0,0,1) scattering plane in \textbf{(a-f)}. Blue lines are fits by Gaussian functions to the data on top of a background (in grey). The gray dots correspond the SF background (Bgr) deduced from XYZ-PA, and then fitted by the gray lines.  On both instruments,  the data have been obtained during two different experimental runs where different sample environments have been used which explains different levels of background. As for all figures showing raw data, the counting time of the reported intensity is 5 minutes/point on Thales  and 15 minutes/point on 4F1.
 Error bars represent one standard deviation. 
 }
\end{figure}

Figure \ref{Fig2} shows momentum scans along the H-direction in the SF channels around Q=(0.5,0,0) in the YBCO69 sample. On both instruments, a peak is observed at low temperature in the superconducting state around H=0.5 (Fig. \ref{Fig2}.a,d). The polarization analysis further proves the magnetic origin of the signal as the SF peak intensity is reduced when one  turns the polarization from $\bf X$ to $\bf Y$ and $\bf Z$. The magnetic signal can be described by a Gaussian profile with a finite momentum distribution, broader than the instrumental resolution.  After deconvolution, its intrinsic linewidth - half width at half maximum (HWHM) - is $\Delta_H$=$0.02\pm0.005$ r.l.u. This corresponds to a finite correlation length along the $a$-axis of  $\xi_{a}=\frac{a}{2\pi \Delta_H}$= $\sim30\pm3\ \text{\AA}$, corresponding to $\sim 7-8$ planar unit-cells.  These measurements were performed using two instrument configurations (different neutron wavevectors) indicating that the signal is static within the lowest energy resolution of Thales of $\sim$ 0.2 meV, implying a characteristic time-scale necessarily much longer than 0.02 ns. This timescale is consistent with the one extracted from muon spin spectroscopy, which indicates slow magnetic fluctuations in the pseudogap state of about 25 ns at this doping level \cite{zhang2018discovery}. Upon increasing temperature, the  short range magnetic peak at  {\bf q}=1/2 is still present in the normal state, in the 100-150K range (Fig. \ref{Fig2}.b,e). However, the peak is significantly reduced at 300K  (Fig. \ref{Fig2}.c,f) being hardly visible. 

\begin{figure}
 \begin{centering}  \includegraphics[width=13cm]{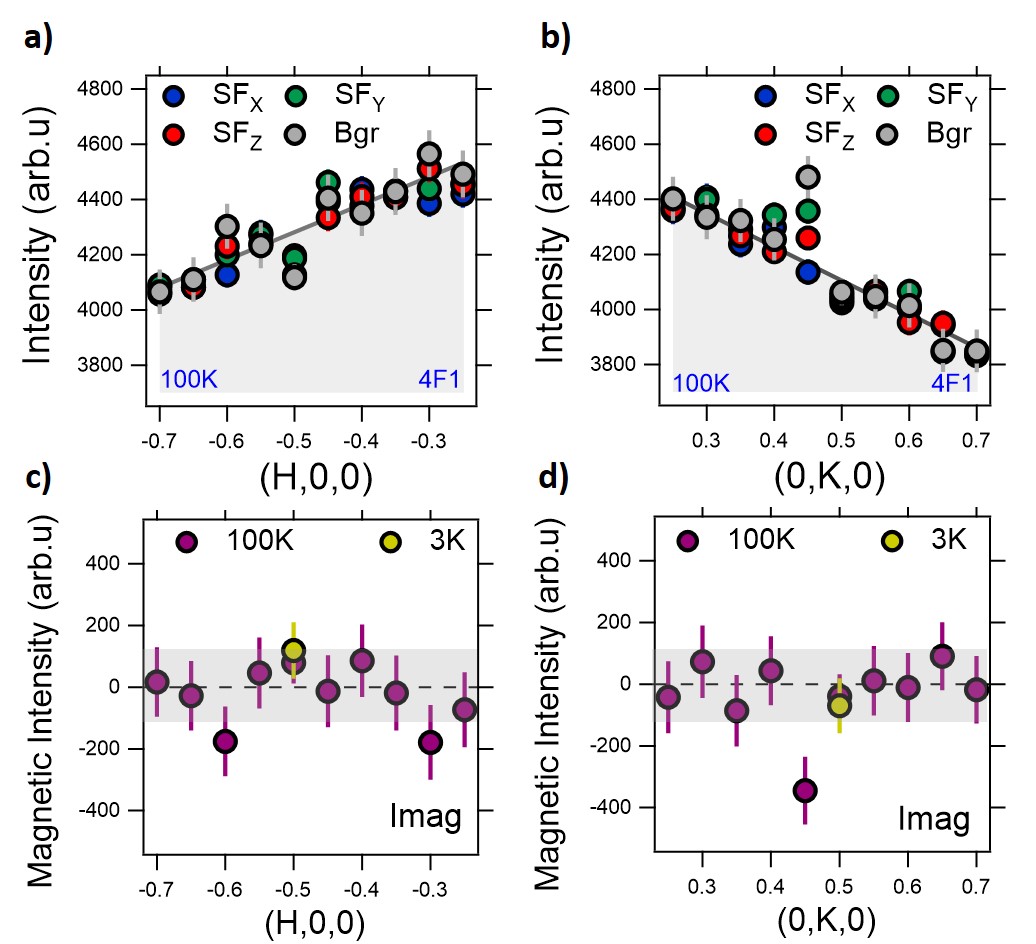} 
 \par\end{centering}
 \clearpage
\caption{\label{Fig3} \textbf{ Absence of {\bf q}=1/2 orbital magnetism in detwinned $\rm YBa_{2}Cu_{3}O_{7}$}:  \textbf{(a-b)} Scans in the spin-flip channel $SF_{X,Y,Z}$ along  \textbf{(a)} The (H,0,0) direction and  \textbf{(b)} The (0,K,0) direction. The background (Bgr) is extracted from XYZ-PA fitted to the gray lines. Scan along  \textbf{(c)} (H,0,0) and  \textbf{(d)} (0,K,0) of the magnetic intensity, $I_{mag}$, as extracted from XYZ-PA at 3K and 100K. Data in  \textbf{(a-d)} were measured on the 4F1 TAS. The sample was aligned in the (1,0,0)/(0,1,0) scattering plane. Error bars (smaller than the points size) represent one standard deviation. 
 }
\end{figure}

Similar measurements were performed in the second sample, YBCO7. As the sample YBCO7 is detwinned, measurements needed to be performed along both {\bf a*} and {\bf b*} directions. Fig. \ref{Fig3}.a-b represent the H-scan and K-scan at 100K in the SF channel. No peak occurs along both directions. Similarly, XYZ-PA confirms the absence of a peak around H=0.5 at 100K (Fig. \ref{Fig3}.c-d). The polarization analysis performed at 3K shows no sizeable feature as well (Fig. \ref{Fig3}.c-d). At such a high doping, the short range magnetism (if any) at {\bf q}=1/2  falls below our threshold of detection.

 As was discussed in ${\rm YBa_{2}Cu_{3}O_{6.6}}$ \cite{Bounoua22}, a magnetic signal at {\bf q}=1/2 is not compatible with known spin structures in the cuprates. Indeed, the antiferromagnetic spin ordering in cuprates is always found near the planar wavevector  {\bf q}=(1/2,1/2). At higher doping, the reported spin fluctuations are also systematically observed close to that wave vector. That corresponds to strong equal first neighbour antiferromagnetic interactions along both a and b. The {\bf q}=1/2 order actually corresponds to a stripe-like spin order as it is observed in the parent compounds of Fe pnictides superconductors, where the magnetic exchange  along the a axis $J_{1a}$ differs from that along the b axis 
 $J_{1b}$  \cite{Dai15}, a situation never encountered in cuprates. Therefore, as at lower hole-doping \cite{Bounoua22}, our results suggest the existence of an orbital magnetism breaking the lattice symmetry, showing that it occurs over a wide doping range that corresponds to the pseudogap regime.

\subsubsection{\label{Ldep} $L$ dependence}

Next, we discuss the $L$-dependence of the magnetic signal in YBCO69 corresponding to the correlations of the  {\bf q}=1/2  magnetism along the c-axis perpendicular to the CuO$_2$ planes. Fig.  \ref{Fig4}.a and b show the neutron intensities at 150K and 280K, respectively. At 150K, two peaks occur with distinct maxima at $L$=0 and $L$=0.5. Both peaks display similar momentum widths, typically about $\Delta_L$=$0.2\pm0.05$ r.l.u. at 150K.   Contrary to the ${\rm YBa_{2}Cu_{3}O_{6.6}}$ sample \cite{Bounoua22}, we cannot prove here that the magnetism reported here is biaxial as the YBCO69 sample is twinned. The existence of two peaks along c contrasts with the single broad peak measured along c in twinned ${\rm YBa_{2}Cu_{3}O_{6.6}}$ at 80K \cite{Bounoua22}. Considering that the peaks of Fig. \ref{Fig4}.a  exhibit c-axis correlation lengths that are different from the one found in the  twinned ${\rm YBa_{2}Cu_{3}O_{6.6}}$ sample,  one can speculate that  the arrangement of magnetic CuO$_2$ layers  along c could be modified by a sample dependent disorder (defects, inhomogeneities, ...). This $L$-dependence also differs from the ones measured in the detwinned  ${\rm YBa_{2}Cu_{3}O_{6.6}}$  sample \cite{Bounoua22}  where a sharp peak was only observed at $L$=0 without any intensity at (0.5,0,0.5) at 5K. 
Note that although the existence of a magnetic signal at (0,0.5,0) was reported\cite{Bounoua22}, the corresponding information at (0,0.5,0.5) was still missing.
This disparity pinpoints a more complex behaviour than  thought at first that could be related to the twinning of our YBCO69 sample.  Indeed, if the signal only occurs at {\bf Q$_b$}=(0,0.5,0.5) but not at {\bf Q$_a$}=(0.5,0,0.5) in a detwinned sample,  the twining may lead to the observed scattering at (0.5,0,0.5). The temperature dependence of the  $L$=0.5 signal may then be different at Q$_a$ and Q$_b$; an information that was left unnoticed in our previous study.  At 280K, the intensity of both peaks along c strongly decreases (Fig.  \ref{Fig4}.b). This shows that the magnetic signal is nearly vanishing near room temperature in agreement with the momentum scan at  L=0  (Fig. \ref{Fig2}.f), and in line with previous results in ${\rm YBa_{2}Cu_{3}O_{6.6}}$  \cite{Bounoua22}. 

One valid question about the c-axis magnetic correlations in YBCO69 is whether they are related to the CuO chains or to the CuO$_2$ planes. In our previous investigation of ${\rm YBa_{2}Cu_{3}O_{6.6}}$ \cite{Bounoua22}, we could resolve that issue as we studied two different samples with different oxygen ordering in the CuO chains that allowed demonstrating that the {\bf q}=1/2 magnetism was not bound to the oxygen superstructure. Additionally, the observed magnetism in the detwinned sample was bi-axial. The  {\bf q}=1/2 magnetism could thus  only originate from the CuO$_2$ planes. Here, the sample structure is Ortho-I with no oxygen between copper spins along $a$ that precludes a magnetic response from the chains at {\bf q}=1/2 along that direction. However, magnetic correlations running only along the $b$ direction could also be  considered. Indeed, quasi-1D fluctuations were reported in ${\rm YBa_{2}Cu_{3}O_{6.93}}$ near optimal doping. They are characterized by a net incommensurate modulation \cite{Mook96} that agrees with band structure calculations associated to nesting effects at 2${\bf k}_F$  of the CuO chains Fermi surface. Related magnetic dynamical correlations around $K\simeq0.7-0.75$ were tentatively reported \cite{Kulda99,Gavrilov01} but do not correspond to our findings. Hence, here again, one can ascribe the origin of the {\bf q}=1/2 magnetism to the CuO$_2$ planes. 

 The scattered magnetic intensity of the {\bf q}=1/2 magnetism at {\bf Q}=(H,0,L), $I_{mag}$, includes the neutron orientation factors related to each magnetic component. It reads:

\begin{equation}
I_{mag} = r_{0}^{2}\,  \,f(Q)^{2}\,  \left[
{\bf m}_b({\bf Q})^2+ (1-|2\pi\frac{H/a}{\bf Q}|^2) {\bf m}_{a}({\bf Q})^2 + (1-|2\pi\frac{L/c}{\bf Q}|^2)  {\bf m}_c({\bf Q})^2  \right]
\label{eq:Full equation}
\end{equation}

where $r_{0}$ stands for the neutron magnetic scattering length, $r_{0}^{2}=290$ mbarn, $f(Q)$ is the magnetic form factor.  $|{\bf Q}|= 2\pi\sqrt{ (H/a)^2 + (L/c)^2}$ is the wave vector modulus expressed  in \AA$^{-1}$ and $a$=3.82 \AA\ and $c$=11.7 \AA\ are the lattice parameters along {\bf a} and {\bf c}, respectively.  ${\bf m}_a({\bf Q})$, ${\bf m}_b({\bf Q})$ and ${\bf m}_c({\bf Q})$ correspond to the Fourier components of magnetic ordered moments along the three crystal directions.  In general,  the {\bf Q}-dependence of the magnetic structure factor relates to the specific magnetic arrangements within the unit cell, in particular here within the CuO$_2$ bilayers on which we do not have sufficient experimental information. To reduce the amount of notations, we omit the {\bf Q} dependence  of the magnetic components hereafter.

 As the YBCO69 sample is twinned, the $(H,0,L)$ scattering plane is indistinguishable from the $(0,K,L)$ scattering plane.  The contributions coming from each domain superpose in the measurements.  Further, as the in-plane components along $a$ and $b$ cannot be distinguished, it is generally assumed that ${\bf m}_a = {\bf m}_b$. We here rather define an in-plane magnetic moment component, ${\bf m}_{ab}$, as ${\bf m}_{ab}^2= {\bf m}_a^2+{\bf m}_b^2$. As a result of these considerations, Eq. \ref{eq:Full equation} can be re-written as,

\begin{equation}
I_{mag} = r_{0}^{2}\,  \,f(Q)^{2}\, \left[ \frac{1}{2} (1+|2\pi\frac{L/c}{{\bf Q}}|^2)  {\bf m}_{ab}^2 + (1-|2\pi\frac{L/c}{\bf Q}|^2)  {\bf m}_c^2  \right]
\label{eq:Full equat}
\end{equation}

 In the spin-flip channel, Eq. \ref{eq:Full equat} corresponds to the full intensity measured for a neutron polarization along X. Fig. \ref{Fig4}.c shows the total scattered magnetic intensity $I_{mag}$ deduced by XYZ polarization analysis (Eq. \ref{Imag}, see ref. \cite{Bounoua20,Bounoua22} for more details). Eq. \ref{eq:Full equat}  includes the neutron orientation factor for each magnetic component as  $(1-|2\pi\frac{L/c}{\bf Q}|^2)$ for the out-of-plane (${\bf m}_c$)
  component and $(1+|2\pi\frac{L/c}{\bf Q}|^2)$ for the in-plane  (${\bf m}_{ab}$) component.  Further, to determine separately the different magnetic components, one needs to apply the polarization analysis. Following our previous reports \cite{Bounoua20,Bounoua22},  the difference between the measured intensities for different polarizations are related to the magnetic components as:

\begin{equation}
\begin{cases}
 {SF}_{\bf X} - {SF}_{\bf Y} = &  r_{0}^{2}\, f(Q)^{2}\, {1\over2} {\bf m}_{ab}^2 \\
 {SF}_{\bf X} - {SF}_{\bf Z} = &   r_{0}^{2}\, f(Q)^{2}\, [ (1-|2\pi\frac{L/c}{\bf Q}|^2)   {\bf m}_c^2+ |2\pi\frac{L/c}{{\bf Q}}|^2 {1\over2} {\bf m}_{ab}^2  ]
\end{cases} \label{XYZ}
\end{equation}

 Next, one defines the quantities $I_c$ and $I_{ab}$ as:  $I_c=(1-|2\pi\frac{L/c}{\bf Q}|^2) f(Q)^{2} {\bf m}_c^2 $, $I_{ab}= \frac{1}{2} (1+|2\pi\frac{L/c}{{\bf Q}}|^2) f(Q)^{2} {\bf m}_{ab}^2$.  In the absence of magnetic chirality, the polarization sum rule holds: $I_{mag}= I_{ab} + I_{c}$ \cite{Bourges11}.  Using XYZ-PA (Eq. \ref{XYZ}), one determines the magnetic in-plane ($I_{ab}$) and out-of-plane ($I_c$) components.  In addition to the results at 150K shown in Fig. \ref{Fig4}.a, we studied the polarization dependence of both peaks over a wide temperature range (4-150K) and averaged out the data to improve statistics  since no significant variation of the ratio $\frac{I_{ab}}{I_c}$ was noticed in temperature. $I_{mag}$,   $I_{ab}$ and $I_c$ are represented in Fig. \ref{Fig4}.c  in units of $\frac{2\pi}{c}L$. At $L$=0, the magnetic signal indicates a predominant component along the c-axis as in the underdoped sample \cite{Bounoua22}.  However, the ratio between in-plane and out-of-plane components, $\frac{I_{ab}}{I_c}$, is larger than for ${\rm YBa_{2}Cu_{3}O_{6.6}}$. The result at $L$=0 contrasts with the polarization analysis of the second peak at $L$=1/2 with a much weaker out-of-plane contribution, resulting in a dominant in-plane component. This actually supports a non-collinear alignment of moments (pointing predominantly along c), which gives rise to distinct magnetic structure factors at $L$=0 and $L$=1/2. 

At larger $L$, the polarization analysis shows that the magnetic signal vanishes rapidly.  On Fig. \ref{Fig4}.c, we report Eq. \ref{eq:Full equat} where  the magnetic form factor is the one expected for a spin moment either of Cu or oxygen atoms (dashed and dotted lines, respectively) normalized to 1 at $L$=0. Clearly, this does not describe our data as that has been previously shown for the {\bf q}=0 IUC magnetic intensity that also decreases rapidly at large $L$ \cite{deAlmeida12}. Beyond the weak $L$ dependence coming from the orientation factors, the fast decay of the magnetic intensity cannot be accounted for by the Cu and/or oxygen spin form factors. This suggests an orbital origin  for the measured magnetism. Along this line, we calculate the measured magnetic intensity in Eq. \ref{eq:Full equat} using the phenomenological form factor along $L$, $|f(L)|^2$, found  for the  {\bf q}=0 IUC  magnetism in 4 distinct cuprate families \cite{deAlmeida12}.  In Fig. \ref{Fig4}.c, $|f(L)|^2$ also normalized to 1 at $L$=0 describes better  
the measured $L$-dependence of the {\bf q}=1/2 magnetic intensity.  In Fig. \ref{Fig4}.c, the full lines represent the product $|f(L)|^2 (1-|2\pi\frac{L/c}{\bf Q}|^2)  {\bf m}_c^2$ for the out-of-plane component or the product $|f(L)|^2 \frac{1}{2}(1+|2\pi\frac{L/c}{\bf Q}|^2)  {\bf m}_{ab}^2$ for the in-plane component, respectively. 
 At variance to the spin form factors,  these relations with $f(L)$ match the characteristic decrease of the {\bf q}=1/2 magnetism for each magnetic component, as it was the case for the {\bf q}=0 IUC magnetism \cite{deAlmeida12}. One can nevertheless notice that the measured out-of-plane intensity is slightly lower than the expectation whereas the opposite trend is found for the in-plane component. That may suggest a different form factor for both components as it has been proposed for loop currents \cite{He12}. On general grounds, it is worth stressing out that the orbital magnetic form factor can be very anisotropic for the case of loop currents due to their extended geometry, delocalized over the Cu and O sites. The fact that both  {\bf q}=0 and  {\bf q}=1/2 magnetisms exhibit the same fast decay along $L$ strongly supports a common orbital origin. 

\subsubsection{\label{intensity} Intensity in absolute units}

Next, we estimate the intensity of the {\bf q}=1/2 magnetism measured in the different samples. To get the magnetic signal in absolute units versus doping, we need to calibrate the magnetic signal by comparison with Bragg peak intensities. We follow the method presented in our previous study in ${\rm YBa_{2}Cu_{3}O_{6.6}}$ \cite{Bounoua22} where the reported magnetic signal at $L$=0 is $q$-integrated in the [a,b] plane. Its hole doping dependence is shown in Fig.  \ref{Fig4}.d. Typically, the amplitude at {\bf q}=1/2 is about three times weaker in the YBCO69 sample compared to ${\rm YBa_{2}Cu_{3}O_{6.6}}$ at lower doping. One clearly sees that the signal at {\bf q}=1/2 decreases upon increasing doping. Interestingly, this decay matches that of the {\bf q}=0 IUC  magnetic scattering as it scales with the magnetic intensity at the Bragg peak {\bf Q}=(1,0,1) reported in   the supplementary information of ref. \cite{Bourges18}. The signal at (0.5,0,0) is  about 5 times weaker than the magnetic intensity at  {\bf Q}=(1,0,1) by comparing both vertical scales in Fig.  \ref{Fig4}.d. It is then typically 10-15 weaker than the magnetic intensity at  {\bf Q}=(1,0,0) \cite{Bounoua22}.  The magnetic moment evolves as the squared root of the neutron intensity. Therefore, using our previous estimation of the magnetic moment in ${\rm YBa_{2}Cu_{3}O_{6.6}}$ \cite{Bounoua22}, a magnetic moment per loop current triangle of
$m_{tr}\sim$ 0.015 $\mu_B$ is found using the assumption that all unit cells contribute uniformly to the magnetic intensity. We stress out that this assumption is  misleading as the {\bf q}=1/2 magnetism is at short-range \cite{Bounoua22}. 

\subsubsection{\label{Tdep} Temperature dependence}

The temperature dependence of the magnetic intensity is further represented in Fig. \ref{Fig5}. The raw neutron intensity at $L$=0 (Fig. \ref{Fig5}.a) decreases upon warming up faster than the background determined by the polarization analysis. Using the XYZ-PA, one can also decompose the signal at $L$=0 into the two different magnetic components as above. The background subtracted total magnetic intensity, $I_{mag}$ is depicted in Fig. \ref{Fig5}.b and the temperature dependencies of the out-of-plane, $I_c$, and in-plane, $I_{ab}$, magnetic components in Fig. \ref{Fig5}.c and d, respectively.  As for ${\rm YBa_{2}Cu_{3}O_{6.6}}$ samples \cite{Bounoua22}, the in-plane component at $L$=0 (Fig. \ref{Fig5}.c) exhibits no change in temperature whereas the out-of-plane component (Fig. \ref{Fig5}.d) captures the whole decrease in temperature where the signal disappears above $\sim$ 250K. The temperature dependence of the background subtracted magnetic intensity at $L$=1/2 is shown in Fig.  \ref{Fig5}.e in comparison with the magnetic intensity obtained by polarization analysis. Due to the weaker amplitude at $L$=1/2, the decomposition of the magnetic signal into the components cannot be performed with confidence. One can nevertheless use these data to determine the temperature where the magnetic response sets in, which is found to be about $T_{mag} \sim 200 K$  where the intensity at $L$=0.5  drops. This temperature is reported in the phase diagram (Fig. \ref{Fig4}.e).  As in ${\rm YBa_{2}Cu_{3}O_{6.85}}$, $T_{mag}$ occurs at higher temperature than the appearance of the CDW that are hardly visible using resonant elastic X-ray scattering in ${\rm YBa_{2}Cu_{3}O_{6+x}}$  above a doping of $\sim$ 0.17 \cite{BlancoCanosa14}.

 The {\bf q}=0 IUC magnetism was not measured in the YBCO69 sample but our data can be compared to the {\bf q}=0 magnetism reported in ${\rm YBa_{2}Cu_{3}O_{6.85}}$ \cite{Mangin15} that had a close hole doping level. We then compare the evolution of temperature dependencies of the {\bf q}=1/2 magnetism with the one reported at {\bf q}=0 in the nearly optimally doped ${\rm YBa_{2}Cu_{3}O_{6.85}}$ \cite{Mangin15}. Interestingly, the temperature dependencies of the out-of-plane magnetic components at (0.5,0,0), in YBCO69, and at {\bf q}=0, in ${\rm YBa_{2}Cu_{3}O_{6.85}}$, agree remarkably (Fig. \ref{Fig5}.d).   $T_{mag}$ corresponds to the inflexion point of the temperature dependence of the out-of-plane component in Fig. \ref{Fig5}.d.
By contrast, the  in-plane component at {\bf q}=0  (Fig. \ref{Fig5}.f) compares well with the temperature evolution at (0.5,0,0.5) where the magnetic signal is predominantly in-plane in character according to the polarization analysis (Fig.\ref{Fig4}.c).  However,  in this comparison, an additional magnetic component at (0.5,0,0.5) is present at all temperature similar to the one found at $L$=0  (Fig. \ref{Fig5}.c). Although the amplitude at  {\bf q}=1/2 is much weaker,  the temperature dependences of the in-plane ($L$=0.5) and out-of-plane ($L$=0) magnetic responses display striking similarities with the {\bf q}=0 IUC response \cite{Mangin15}.  There is however a noticeable difference. At {\bf q}=0, both components occur at the same $L$ values yielding an apparent tilt of the magnetic moment \cite{Tang18,bourges2021-online}, as expected for collinear and canted magnetic moments.  At {\bf q}=1/2, the components occur at two different $L$ positions suggesting a more complex stacking of the magnetic correlations along the c-axis,   more consistent with a non-collinear  arrangement  of canted magnetic moments. Typically, the alternating in-plane component is related to a doubling of the unit-cell along c, which does not occur for the {\bf q}=0 IUC magnetism. 
As the doubling of the unit cell along c affects the magnetism of 4 CuO$_2$ layers ({\it i.e.} 2 bilayers), an in-plane magnetic moment turning at 90$^\circ$ from one plane to the next along c could for instance produce such a doubling of the periodicity along c.

 The phase diagram of Fig. \ref{Fig4}.e shows the universal character of the {\bf q}=1/2  magnetism. It corresponds to the region in doping associated with the existence of the pseudogap in YBCO. The ordering temperature for both the long range  {\bf q}=0 IUC magnetism and the shorter range {\bf q}=1/2 magnetism decreases first linearly in the underdoped region, then flattens near optimal doping, and vanishes beyond, concomitantly with the PG. 

\section{\label{discussion} Discussion and concluding remarks}

Before going into a detailed interpretation of the data in terms of orbital moments, it is important to dismiss extrinsic contributions, for instance due to magnetic peak in cupric oxide (CuO) impurities. Indeed,  CuO is antiferromagnetic (AF) \cite{Forsyth,Ain92} with an ordered moment of 0.65 $\mu_B$ and T$_N=$230 K, its strongest (1/2,0,-1/2) AF peak has a $|Q|$ very close to the (0.5,0,0.5) momentum of YBCO69.  The same issue was discussed and dismissed in our previous report of $q=1/2$ magnetism \cite{Bounoua22}. As we discuss extensively in section 2 of the Supplemental Material  \cite{SuppMat}, CuO powder and crystallites impurities can be ruled as well in our YBCO69 sample as the origin of the reported magnetic peaks.  In particular, to be able to explain the neutron data, the CuO impurities have  to be nano-size grains of $\sim$ 25 \AA. As discussed in ref.  \cite{nanoscaleCuO},  the N\' eel temperature in such nano-size grains of CuO is below 50K which is inconsistent with our reported temperature dependence.

As discussed in our previous report \cite{Bounoua22}, these results can be described by an hidden magnetic texture mixing small regions in the CuO$_2$ plane having loop currents that break the lattice translational symmetry with larger regions in the CuO$_2$ plane where the lattice translation symmetry is preserved. One can then build a real space picture of 30x30 unit cells paved by anapoles representing loop current states (represented in Fig. \ref{Fig6}.a), with a bubble of loop currents breaking the lattice translational symmetry ("2x2" phase) connecting larger domains of the uniform loop currents phase as it was theoretically proposed \cite{Varma19}. The Fig. \ref{Fig6}.b shows a realistic real-space example of such a 2D pattern where the arrows represent the anapole polar vectors in each unit cell. The central pattern displays a correlation length of 7-8 units cells consistent with our measurements. One can then calculate the structure factor of this pattern in reciprocal space. The Fig. \ref{Fig6}.c shows in log-scale such a structure factor convoluted by the neutron 2D momentum resolution (FWHM=$\Delta_H$=$\Delta_K$=0.05 r.l.u. ) corresponding to our measurements on 4F1. Two main features are found. First, the strongest magnetic peaks are the ones at the Bragg positions (1,0) and (0,1) of the structure corresponding to the observed {\bf q}=0 IUC magnetism \cite{Fauque06,Mook08,Baledent11,Mangin15,Tang18,bourges2021-online}. Due to the supercell size of 60 units cells of Fig. \ref{Fig6}.a, a set of satellites magnetic peaks located at $\pm{1\over{60}}$ from the atomic Bragg peak positions should show up in the structure factor. However, they are blurred by the instrumental momentum resolution to give rise to an elongated shape in the longitudinal direction. Second, a diffuse scattering is observed at both (0.5,0) and (0,0.5) corresponding to the {\bf q}=1/2 magnetism. The size of the "2x2" bubble over the total size of the pattern determines the ratio of intensities between both features. Finally, our polarized neutron diffraction data qualitatively support the existence of a complex magnetic pattern. However, the accuracy of the data does not allow one to pinpoint the exact magnetic pattern and a multitude of other real space patterns can be considered to give rise to an intensity distribution in momentum space similar to Fig. \ref{Fig6}.c.  The picture discussed in Fig. \ref{Fig6}.b  is just an attempt to give a coherent picture of all the neutron diffraction measurements done so far. Interestingly, that picture got very recently some momentum by the observation of a topological magnetic texture in the PG state of YBCO by Lorentz transmition electronic microscope (LTEM) \cite{LTEM}. A vortex-like magnetization density was reported in the CuO$_2$ sheets, with a magnetic texture of total size of about 100 nm. Its intensity distribution shows a maximum around 25 nm,  corresponding roughly to 60 unit cells. It is tempting to relate this magnetic texture to large domains such as the one represented here in Fig. \ref{Fig6}.b. More work is necessary to quantify the exact relation. 

To conclude, our polarized neutron scattering measurements on a high quality single crystal of ${\rm YBa_{2}Cu_{3}O_{6.9}}$ generalize the existence of magnetic correlations at {\bf q}=1/2 that break the lattice translational symmetry, over a wide hole doping range matching the pseudogap region of cuprate superconductors.  Considering the location of the {\bf q}=1/2 magnetism in momentum space and the $L$-dependence of its amplitude, decreasing at large $L$, this magnetism is likely to be orbital in origin. Its amplitude is further reduced at larger doping near optimal doping compared to underdoped  ${\rm YBa_{2}Cu_{3}O_{6.6}}$ cuprates  \cite{Bounoua22}.  It is interesting to notice that the size of the {\bf q}=1/2 magnetism does not change significantly with doping with similar measured correlations lengths within the CuO$_2$ planes.  In addition, we find here a dichotomy of the components of the magnetic response along the c-axis. The out-of-plane component, naturally associated with loop currents circulating within the CuO$_2$ plane, respects the c-axis stacking whereas the in-plane component is doubling the unit cell. This is surprising as this dichotomy is not observed for the  {\bf q}=0 IUC magnetic response, yielding an apparent tilt of the magnetic moment \cite{Fauque06,Mook08,Baledent11,Mangin15,Tang18,bourges2021-online}, and deserves further investigations.  

\bibliography{papier-ok}
\noindent \textbf{Acknowledgments}

We thank F. Maignen, C. Meunier and S. Klimko for their valuable technical assistance and installation of MuPaD on 4F1. We thank 
the group Synth\`ese Propri\'et\'es et Mod\'elisation des Mat\'eriaux de l'Institut de Chimie Mol\'eculaire et des Mat\'eriaux d'Orsay de l'Universit\'e Paris-Saclay for the sample preparation and characterization. The data obtained on Thales at ILL are available at https://doi.org/10.5291/ILL-DATA.4-02-600. We acknowledge supports from the project NirvAna (contract ANR-14-OHRI-0010)  of the French Agence Nationale de la Recherche (ANR) and from the GenLoop  project  of the LabEX PALM (contract ANR-10-LABX-0039-PALM). 

Corresponding authors: Dalila Bounoua or Philippe Bourges.
\section*{Ethics declarations}
\subsection*{Competing interests}
The authors declare no competing interests.

\begin{figure} [b!]
   \begin{centering}
  \includegraphics[width=15cm]{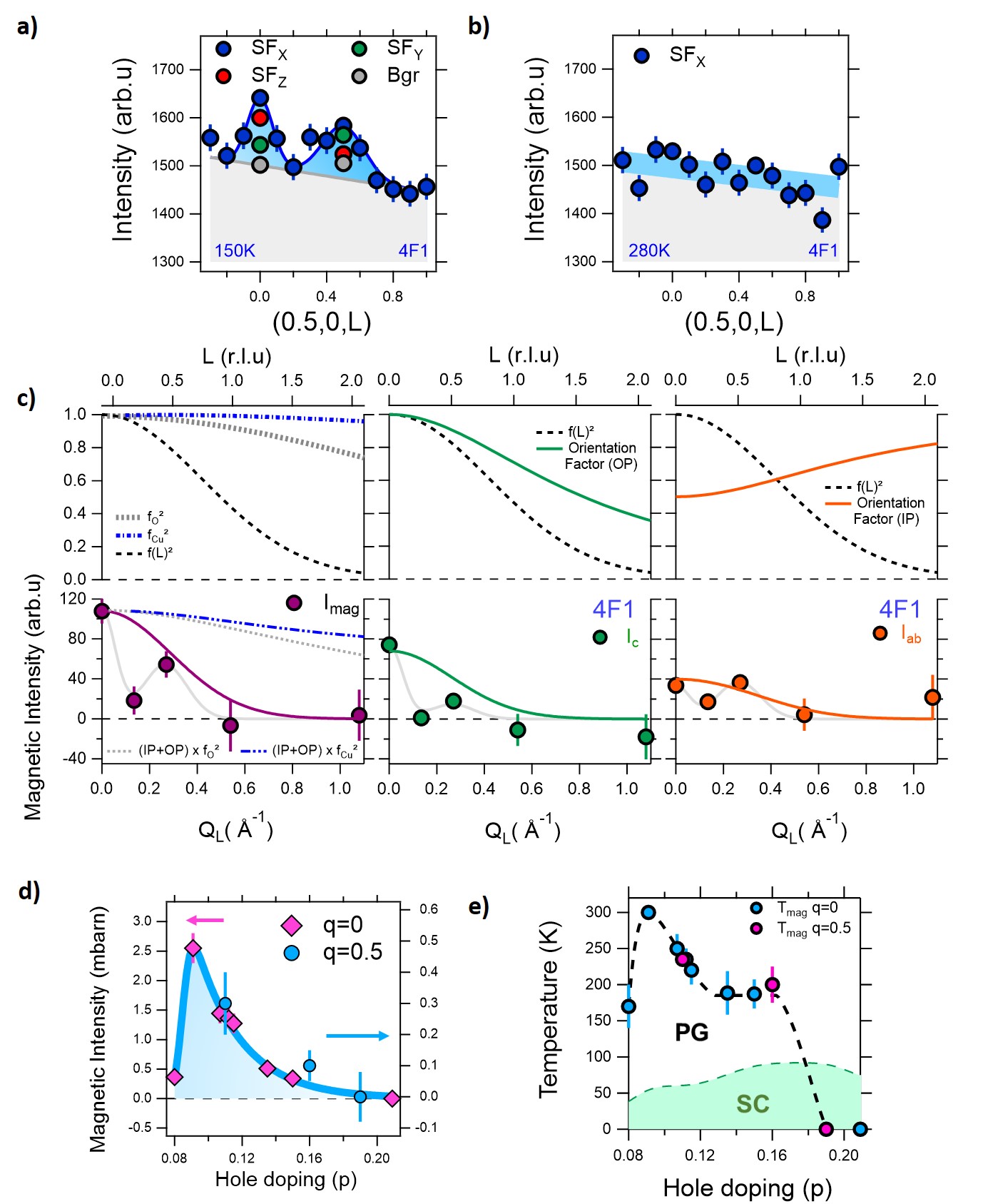} 
 \par\end{centering}
  \caption{(Caption next page.)}
\end{figure}
\addtocounter{figure}{-1}
\begin{figure} [t!]
 \caption{\label{Fig4} \textbf{$L$-dependence of the {\bf q}=1/2 orbital magnetism in $\rm YBa_{2}Cu_{3}O_{6.9}$}:
\textbf{(a-b)} L-Scans along (0.5,0,L) in the spin-flip channel $SF_{X,Y,Z}$ collected at \textbf{(a)} 150K and \textbf{(b)} 280K. The magnetic intensity in  \textbf{(a)}  appears as a Gaussian signal centered at Q=(0.5,0,0) and Q=(0.5,0,0.5). Blue lines are fits by Gaussian functions to the data. The background (Bgr) is extracted from XYZ-PA fitted to the gray lines. \textbf{(c)} (lower part) $L$-dependence along (0.5,0,L) of total magnetic scattering (left), out-of plane magnetic contribution $I_c$ (center) and  in-plane magnetic contribution $I_{ab}$ (right), as extracted from XYZ-PA.  The data were measured at various temperatures in the range 5-150K and averaged out.  In the left upper panel, the form factor for spins from Cu and O are shown with the phenomenological form factor,  $|f(L)|^2$ (black dashed line), found in ref. \cite{deAlmeida12} for the {\bf q}=0 magnetism. All form factors are normalized to 1 at $L$=0. The two other upper panels show the orientation factor discussed in the text for each magnetic component. In the lower panels, the full lines describing $I_c$ (center) and $I_{ab}$ (right) are obtained from the magnetic orientation factor multiplied to the form factor $|f(L)|^2$ (see text). The full line in the left panel is a fit of Eq. \ref{eq:Full equat}. The light grey curves are guide to the eye. Data in \textbf{(a-c)} were measured on the 4F1 TAS with the sample aligned in the (1,0,0)/(0,0,1) scattering plane.  \textbf{(d)} Comparison of the scattered intensity in absolute units (mbarn) versus doping observed for the {\bf q}=1/2 magnetism with the {\bf q}=0 signal measured  in $\rm YBa_{2}Cu_{3}O_{6+x}$ \cite{Fauque06,Mook08,Baledent11,Mangin15,Bourges18,Bounoua22}. The line is a guide to the eye. \textbf{(e)}  Summary phase diagram of the observed onset temperature, T$_{mag}$, for the  {\bf q}=0 and {\bf q}=0.5 magnetism versus doping \cite{Bourges18,Bounoua22}.  }
\end{figure}

\begin{figure} [b!]
   \begin{centering}
  \includegraphics[width=15cm]{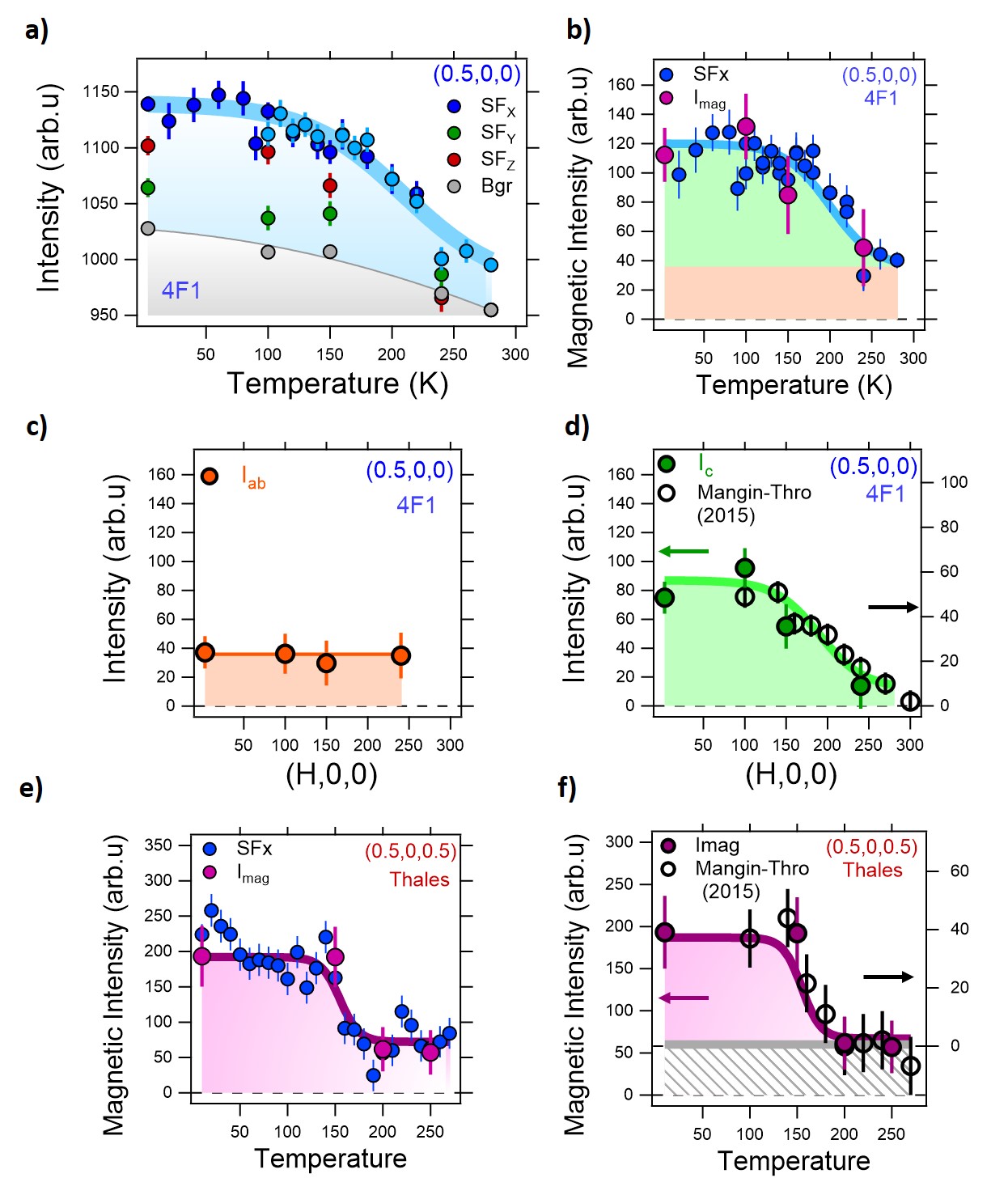} 
 \par\end{centering}
  \caption{(Caption next page.)}
\end{figure}
\addtocounter{figure}{-1}
\begin{figure} [t!]
\caption{\label{Fig5} \textbf{Temperature dependence of the {\bf q}=1/2 orbital magnetism in $\rm YBa_{2}Cu_{3}O_{6.9}$}: \textbf{(a)}  Temperature dependence of the magnetic signal at {\bf Q}=(0.5,0,0) measured in the $SF_{X,Y,Z}$ channels and the corresponding sloping background (Bgr) as extracted from XYZ-PA. The dark and light blue points correspond to two distinct datasets. \textbf{(b)} Temperature dependence of the background subtracted magnetic scattering at {\bf Q}=(0.5,0,0):  $SF_X$-Bgr from panel  \textbf{(a)} for the blue points and as extracted from XYZ-PA for the purple points. The in-plane magnetic response corresponding to $I_{ab}$ (shown independently in \textbf{(c)}) is indicated by the orange area while the out-of-plane magnetic response corresponding to $I_c$ (shown independently in \textbf{(d)}) is indicated by the green area. The blue and violet symbols correspond to the full magnetic scattering ($I_c$ + $I_{ab}$). The black open symbols in  \textbf{(d)} correspond to the temperature dependence of the out-of-plane magnetic response for the {\bf q}=0 magnetism in $\rm YBa_{2}Cu_{3}O_{6.85}$ \cite{Mangin15}. \textbf{(e-f)} Temperature dependence of the magnetic intensity at {\bf Q}=(0.5,0,0.5):  $SF_X$-Bgr of panel for the blue points and as extracted from XYZ-PA for the purple points. \textbf{(f)}  Comparison between the current work with the temperature dependence of the in-plane magnetic response  (open symbols, right scale) for the {\bf q}=0 magnetism in $\rm YBa_{2}Cu_{3}O_{6.85}$ \cite{Mangin15}. The {\bf q}=0 magnetism is shifted by a constant  (dashed grey area)  for the sake of comparison with \cite{Mangin15}. Data in  \textbf{(a-d)} were measured on the 4F1 TAS and data in \textbf{(e-f)} were obtained on the Thales TAS. The sample was aligned in the (1,0,0)/(0,0,1) scattering plane. Error bars represent one standard deviation. Lines are guide to the eye.
 }
\end{figure}

\begin{figure} [b!]
   \begin{centering}
  \includegraphics[width=15cm]{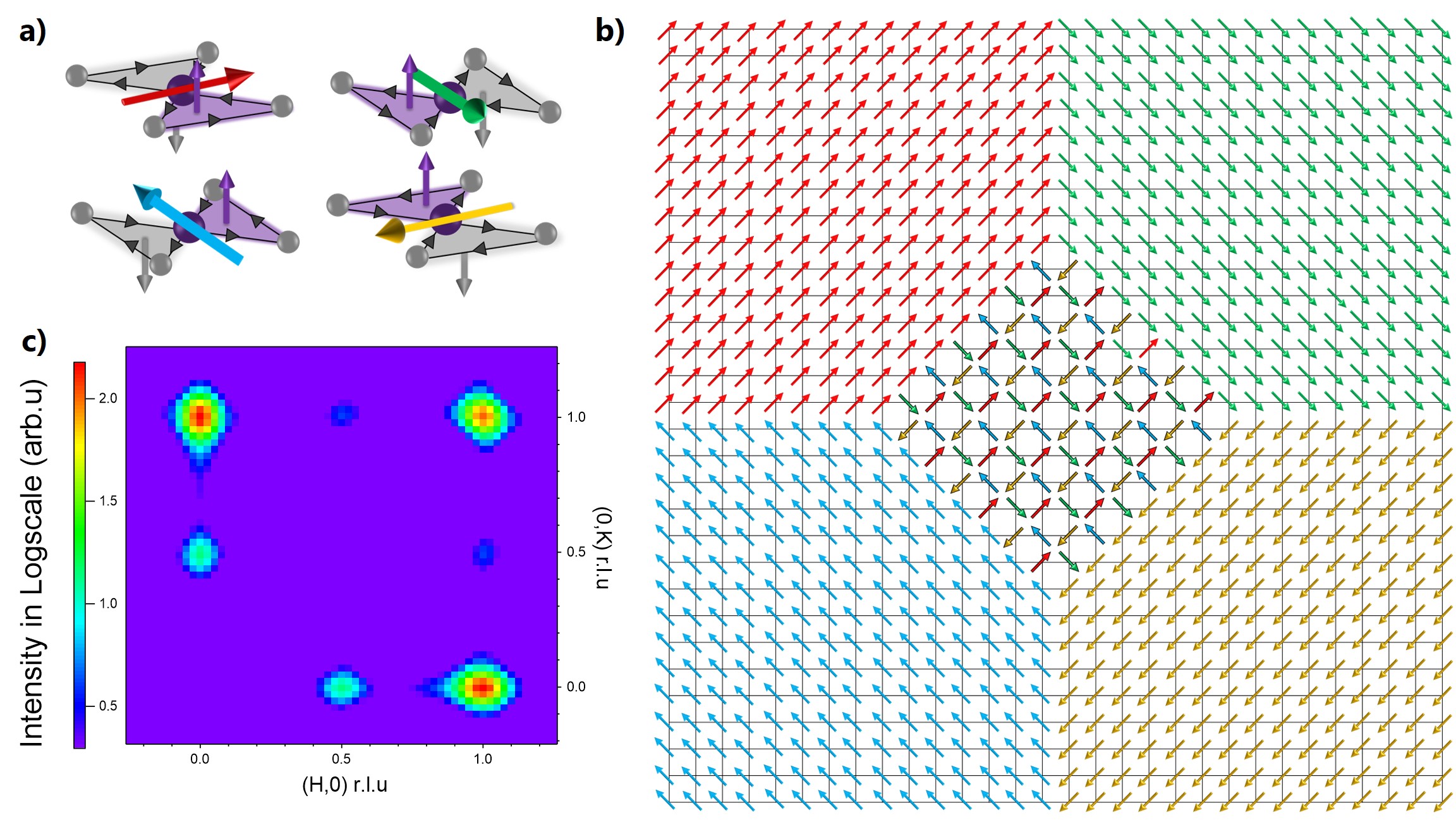} 
 \par\end{centering}
  \caption{(Caption next page.)}
\end{figure}
\addtocounter{figure}{-1}
\begin{figure} [t!]
\caption{\label{Fig6} \textbf{Structure factor calculation for the hidden magnetic structure including  {\bf q}=1/2 and {\bf q}=0 magnetism in $\rm YBa_{2}Cu_{3}O_{6.9}$}: \textbf{(a)} (from \cite{Bounoua22}) The spontaneously circulating LC state on a CuO$_{2}$ unit cell comprising two loops circulating clockwise (in gray) and anti-clockwise (in purple) leading to two magnetic moments perpendicular to the plane. The gray and purple arrows represent magnetic moments along the c axis. The 4 possible patterns are characterized by horizontal colored arrows, corresponding to 4 distinct anapole polar vectors centered at the Cu-site. \textbf{(b)} Proposition of a 2D magnetic texture with 30x30 unit cells paved by the 4 anapoles (LC states) shown in panel \textbf{(a)}. The 4  different anapole orientations correspond to rows with different colors. The central cluster reproduces the {\bf q}=1/2 magnetism with 2x2 LC patterns binding large ferro-anapolar domains corresponding to the IUC magnetism.  \textbf{(c)} Structure factor calculation in log-scale for the LC magnetic  pattern reported in  \textbf{(b)}, convoluted with the instrumental resolution.  The $(H,K)$ intensity map shows magnetic scatterings at (1,0), (0,1) and (1,1) associated with the  {\bf q}=0  magnetism and those at (0.5,0) and (0,0.5)  related to the {\bf q}=1/2 magnetism. 
}
\end{figure}

\newpage

\includepdf[pages=-,scale=.8]{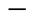}

\end{document}